\documentclass[aps,pra,reprint,superscriptaddress,longbibliography]{revtex4-2}

\usepackage{graphicx} % Including figure files
\usepackage{dcolumn} % Align table columns on decimal point
\usepackage{bm} % bold math
\usepackage{hyperref}% add hypertext capabilities
\usepackage{physics}
\usepackage[dvipsnames]{xcolor}

\begin{document}

\title{Bose-Einstein condensate of ultracold sodium-rubidium molecules with tunable dipolar interactions}

\author{Zhaopeng Shi}
\thanks{These two authors contributed equally to this work.}
\affiliation{Department of Physics, The Chinese University of Hong Kong, Hong Kong SAR, China}
\affiliation{State Key Laboratory of Quantum Information Technologies and Materials, The Chinese University of Hong Kong, Hong Kong SAR, China}

\author{Zerong Huang}
\thanks{These two authors contributed equally to this work.}
\affiliation{Department of Physics, The Chinese University of Hong Kong, Hong Kong SAR, China}
\affiliation{State Key Laboratory of Quantum Information Technologies and Materials, The Chinese University of Hong Kong, Hong Kong SAR, China}

\author{Fulin Deng}
\affiliation{Institute of Theoretical Physics, Chinese Academy of Sciences, Beijing 100190, China}

\author{Wei-Jian Jin}
\affiliation{Institute of Theoretical Physics, Chinese Academy of Sciences, Beijing 100190, China}

\author{Su Yi}
\affiliation{Institute of Fundamental Physics and Quantum Technology and School of Physics Science and Technology, Ningbo University, Ningbo, 315211, China}

\author{Tao Shi}
\email{tshi@itp.ac.cn}
\affiliation{Institute of Theoretical Physics, Chinese Academy of Sciences, Beijing 100190, China}

\author{Dajun Wang}
\email{djwang@cuhk.edu.hk}
\affiliation{Department of Physics, The Chinese University of Hong Kong, Hong Kong SAR, China}
\affiliation{State Key Laboratory of Quantum Information Technologies and Materials, The Chinese University of Hong Kong, Hong Kong SAR, China}

\date{\today}

\begin{abstract}

Realizing Bose-Einstein condensation of polar molecules is a long-standing challenge in ultracold physics and quantum science due to near-universal two-body collisional losses. Here, we report the production of a Bose-Einstein condensate of ground-state sodium-rubidium molecules via high efficiency evaporative cooling, with losses suppressed using the dual microwave shielding technique. The ability to tune the dipolar interaction between these ultracold polar molecules is crucial for producing the condensate and enables exciting prospects for future applications. We explore different regimes of dipolar interactions, realizing both the gas phase and the quantum droplet phase of the molecular condensate. This work opens new avenues for investigating quantum matter with strong dipolar interactions and for quantum simulation of long-range many-body systems.

\end{abstract}

\maketitle

\section*{Introduction}

A Bose-Einstein condensate (BEC) of ultracold polar molecules (UPMs) provides an ideal starting point for exploring their tremendous potential in quantum science~\cite{carr2009cold,bohn2017cold,baranov2012condensed}. Enabled by their unique combination of quantum resources, including rich internal structures, long single particle lifetime, and the long-range, anisotropic electric dipole-dipole interaction (DDI), the possible applications of UPMs span a vast range of topics, including quantum simulation of many-body physics~\cite{buchler2007strongly,yi2007novel,micheli2006toolbox}, quantum computation~\cite{demille2002quantum,cornish2024quantum}, precision measurement~\cite{demille2017probing}, and ultracold chemistry~\cite{hu2019direct,karman2024ultracold}. However, compared to ultracold atoms, these complex resources also make the production of UPMs and the realization of a BEC significantly more challenging. After nearly two decades of worldwide efforts, several production techniques for ground-state UPMs are now well-established~\cite{langen2024quantum}. In particular, the ultracold heteronuclear atom association method has become highly robust, routinely producing samples with phase-space density (PSD) on the percent level~\cite{ni2008high}. However, further evaporative cooling of these samples to quantum degeneracy is challenging due to severe two-body losses. These losses result from the formation of two-molecule complexes, which are rapidly excited by trapping light, with near-universal loss rates dictated by long-range collisional dynamics~\cite{mayle2013scattering, christianen2019photoinduced}.

To overcome the loss problem, significant efforts have been dedicated to developing various loss suppression methods. For fermionic UPMs, both DC electric field~\cite{quemener2011dynamics,avdeenkov2006suppression,matsuda2020resonant} and microwave field shielding~\cite{karman2018microwave,lassabliere2018controlling,anderegg2021observation}, achieved by engineering long-range potential barriers via DDI, have been successfully employed. These techniques have enabled evaporative cooling and the production of quantum degenerate Fermi gases of UPMs~\cite{li2021tuning,schindewolf2022evaporation,valtolina2020dipolar}. 

For bosonic UPMs, which interact via the barrierless s-wave channel, mitigating the two-body losses is more challenging. Microwave shielding using a single circularly polarized microwave field has been implemented to reduce loss rates by a factor of over 100, as demonstrated with $^{23}$Na$^{133}$Cs and $^{23}$Na$^{87}$Rb molecules~\cite{bigagli2023collisionally,lin2023microwave}. While promising, this is still insufficient to support evaporative cooling into the BEC regime. Unfortunately, the single microwave field induces not only a long-range barrier, but also a shallow potential well at even longer range. Further improvement is hindered by three-body losses due to field-linked states inside the well~\cite{chen2023field,stevenson2024three}. A breakthrough in 2024 addressed this issue by introducing a second microwave with linear polarization, enabling fine-tuning of the DDI and thus the potential well~\cite{gorshkov2008suppression,karman2025double}. Using this dual microwave dressing technique, the first BEC of UPMs was finally achieved successfully with $^{23}$Na$^{133}$Cs molecules~\cite{bigagli2024observation}.   

The groundbreaking work of $^{23}$Na$^{133}$Cs paved a viable path toward overcoming the long-standing challenge of creating BECs of UPMs. Extending this achievement to more molecular species is crucial for exploiting their unique properties to unlock the full potential of UPMs. In this work, building on our earlier successes with single microwave loss shielding~\cite{lin2023microwave}, we report the production of a BEC of ground-state $^{23}$Na$^{87}$Rb molecules using dual microwave dressing. Leveraging the control of DDI provided by this technique, we suppress the two-body and three-body losses by tuning the long-range potential well accompanying the repulsive shielding core. This enables highly efficient evaporative cooling, leading to the creation of a BEC of $^{23}$Na$^{87}$Rb molecules. Furthermore, we demonstrate smooth transitions between different dipolar regimes, including the observations of the electrostriction effect in the gas phase and a self-bound $^{23}$Na$^{87}$Rb quantum liquid droplet. Our results establish a versatile platform for exploring dipolar quantum physics with UPMs and validate dual microwave dressing as a general method for creating quantum degenerate samples of UPMs.

\section*{Double microwave dressing}

\begin{figure} 
	\centering
	\includegraphics[width=0.8\linewidth]{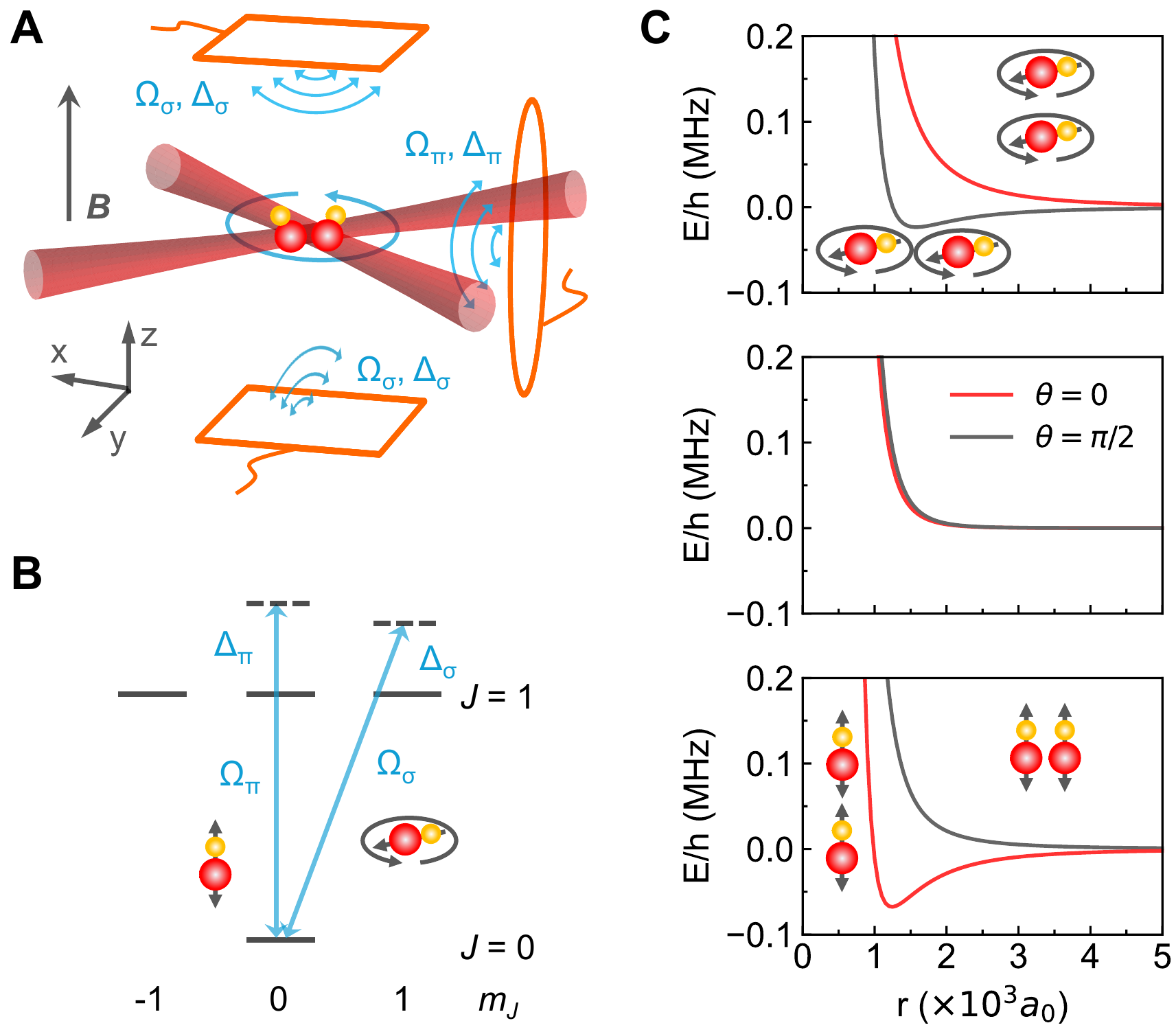}
	\caption{\textbf{Dual microwave dressing for tuning the DDI.} (A) Experimental setup: the $\sigma^+$ polarization microwave is generated by crossed loop antennas placed above and below the optically trapped NaRb sample, and the $\pi$ polarization microwave is from a side loop. The detection is performed along the y-direction. (B) The $\sigma^+$ and $\pi$ microwaves induce rotating and oscillating dipoles, respectively. (C) Due to opposite DDI signs between oscillating versus rotating dipoles, molecule-molecule interactions can be precisely tuned. Middle plot shows complete first-order DDI cancellation with purely repulsive potentials in all directions. Top and bottom plots are partially canceled and over-compensated rotating DDI regimes, respectively, achieved by adjusting only $\Delta_{\pi}$ .     
}
	\label{fig1} 
\end{figure}

In the absence of loss suppression, the two-body loss rate coefficient $\beta_{\rm{in}}$ of $^{23}$Na$^{87}$Rb has a large value of several times $10^{-10}\,\rm{cm^3/s}$~\cite{ye2018collisions,guo2018dipolar}. In single microwave shielding, a circularly polarized ($\sigma^+$) microwave field, with Rabi frequency $\Omega_{\sigma}$ and blue detuning $\Delta_{\sigma}$ relative to the $| J = 0, m_J = 0 \rangle \leftrightarrow | J = 1, m_J = 1 \rangle$ rotational transition [Fig.~\ref{fig1}B], induces a long-range barrier, which can suppress $\beta_{\rm{in}}$ to $3\times 10^{-12}\,\rm{cm^3/s}$~\cite{lin2023microwave}. Here, $J$ is the rotational quantum number, $m_J$ is its projection along the quantization axis defined by the vertical magnetic field. To suppress $\beta_{\rm{in}}$ further, the depth of the shallow potential well must be decreased to the extend that field-linked states cannot be supported. As illustrated in Fig.~\ref{fig1}, this can be achieved by introducing a linearly polarized ($\pi$) microwave field on the $|0, 0 \rangle \leftrightarrow |1, 0 \rangle$ rotational transition, with Rabi frequency $\Omega_{\pi}$ and blue detuning $\Delta_{\pi}$~\cite{karman2025double,bigagli2024observation}. The DDI between the oscillating dipoles induced by the $\pi$ polarization microwave has the opposite sign compared to that between the rotating dipoles induced by the $\sigma^+$ polarization microwave. This allows for fine control of the interaction potential for all dipole orientations by tuning the microwave parameters~\cite{bigagli2024observation,deng2025two,karman2025double,yuan2025extreme}.

Similar to the case of single microwave dressing~\cite{deng2023effective}, the interaction between two molecules in the presence of dual microwave dressing is represented by an effective potential in spherical coordinates as~\cite{deng2025two,note1}

\begin{equation}
	V_{\rm{eff}}(\textbf{r}) = \frac{C_6(\theta)}{r^6} + \frac{C_3}{r^3}(3\cos^2\theta - 1),
	\label{eq1} 
\end{equation}
with $\theta$ the polar angle between $\textbf{r}$ and the quantization axis. The $C_6$ term represents the second-order DDI, which is always repulsive and can be tuned by the changing the microwave parameters. The first-order DDI is governed by 
\begin{equation}
	C_3 = \frac{d_0^2}{24\pi\varepsilon_0}(3\cos 2\beta - 1)\cos^2\alpha \sin^2\alpha,
	\label{eq2} 
\end{equation}  

where $d_0$ is the permanent dipole moment of the molecule. The $C_3$ coefficient can be either positive or negative when the Euler mixing angles $\beta$ and $\alpha$ are tuned by the microwave parameters. In particular, it becomes zero when $\beta$, which measures the contributions of the rotating and oscillating dipoles, is set to $35.3^\circ$. In this case, the molecules experience a purely long-range repulsive shielding core generated only by the $C_6$ term. Since no field-linked states can form, three-body losses are fully suppressed, and lower $\beta_{\rm{in}}$ can be obtained. 

However, as will be discussed later, achieving full cancellation of the first-order DDI requires controlling the two microwave fields with extremely high precision, which is very challenging experimentally. Fortunately, although a non-zero first-order DDI is always accompanied by the long-range potential well, three-body losses should not pose a serious issue as long as the well is not deep enough to support bound states~\cite{stevenson2024three}.

\section*{Experiment platform}

We start the experiment from an optically trapped sample of $8(1) \times10^4$ $^{23}$Na$^{87}$Rb molecules in the ground ro-vibrational level, prepared via magneto-association followed by stimulated Raman adiabatic passage (STIRAP) population transfer~\cite{Guo2016,note1}. The pancake-shaped optical dipole trap (ODT) is formed by crossing two elliptically shaped 1064.2 nm laser beams. The typical initial sample temperature is $T = 730(30)$ nK, and the calculated PSD is 0.008(1). Immediately after STIRAP, we ramp up the $\sigma^+$ polarization microwave field, followed sequentially by the $\pi$ polarization microwave field. The entire microwave ramping process takes 100 $\mu$s, which is slow enough to prepare the molecules adiabatically in the doubly dressed state while minimizing two-body losses from non-shielded collisions.

Microwave shielding suppresses the role of short-range interactions, making the collisional dynamics predominantly governed by the universal DDI~\cite{dutta2025universality}. Consequently, all collisional parameters, including the inelastic rate coefficient $\beta_{\rm{in}}$ and the elastic collision rate coefficient $\beta_{\rm{el}}$, exhibit scaling behaviors dictated by $d_0$ and the molecular mass $m$. Since the $d_0$ of $^{23}$Na$^{87}$Rb is 3.2 D, much less than the 4.75 D dipole moment of $^{23}$Na$^{133}$Cs, achieving a similar level of shielding effects requires microwave Rabi frequencies larger than 10 MHz. 

As depicted in Fig.~\ref{fig1}A, we generate the $\sigma^+$ polarization microwave with two crossed loop antennas and the $\pi$ polarization microwave with a single loop antenna. The polarizations are calibrated and controlled with the help of Rabi spectroscopy at low power and dressed-state spectroscopy at high microwave power. The signals are highly filtered with narrow bandpass filters and a home-built microwave resonant cavity to suppress phase noise before and after the high-power amplifiers. With this system, we obtain maximum Rabi frequencies of 25 MHz and 11.5 MHz for the $\sigma^+$ and $\pi$ polarizations, respectively. The one-body lifetime for Rabi frequencies around 10 MHz, measured with very low-density thermal samples, is 12(2) s. This indicates a phase noise below -160 dBc/Hz for sidebands offset by the Rabi frequencies from the carriers~\cite{lin2023microwave}.

Despite the high degree of control, microwave polarization impurities persist in our system and significantly affect the interaction potential~\cite{note1}. One of the main contributions to these polarization impurities is the misalignment between the two microwaves. Based on the measured ratios of the microwave electric field components of both the $\sigma^+$ and $\pi$ signals at low power, we can only estimate the lower and an upper bounds for the misalignment polar angle as $2.7^\circ \leq \vartheta \leq 9.9^\circ$ (the definition of $\vartheta$ is provided later in the inset of Fig.~\ref{fig4}B). Additionally, the azimuthal angle of the misalignment remains undetermined. This misalignment introduces additional circular polarization components at the frequency of the $\pi$ polarization microwave. 

For detection~\cite{note1}, we first release the sample from the ODT to undergo time-of-flight (TOF) expansion for controlled durations in the presence of the microwave fields, which are essential to suppress loss during the TOF. The microwaves are then ramped down in reverse order to convert the molecules back to their bare state $|0, 0 \rangle$. This is followed by dissociation of the ground-state molecules using a reversed bound-to-free STIRAP process. The molecule number and RMS sizes of the sample are then obtained from the high-magnetic-field absorption images of the fragmented $^{87}$Rb atoms~\cite{jia2020detection}. As the molecule sample falls down during TOF and the Raman beam sizes are small, a galvo is used to dynamically steer the Raman beams to ensure their alignment with the sample. This setup enables probing of the molecules for TOF durations of up to 20 ms, which is necessary to observe clear signatures of the BEC phase transition.

\section*{Evaporative cooling and Bose-Einstein condensation}

\begin{figure*} 
	\centering
	\includegraphics[width=0.8\linewidth]{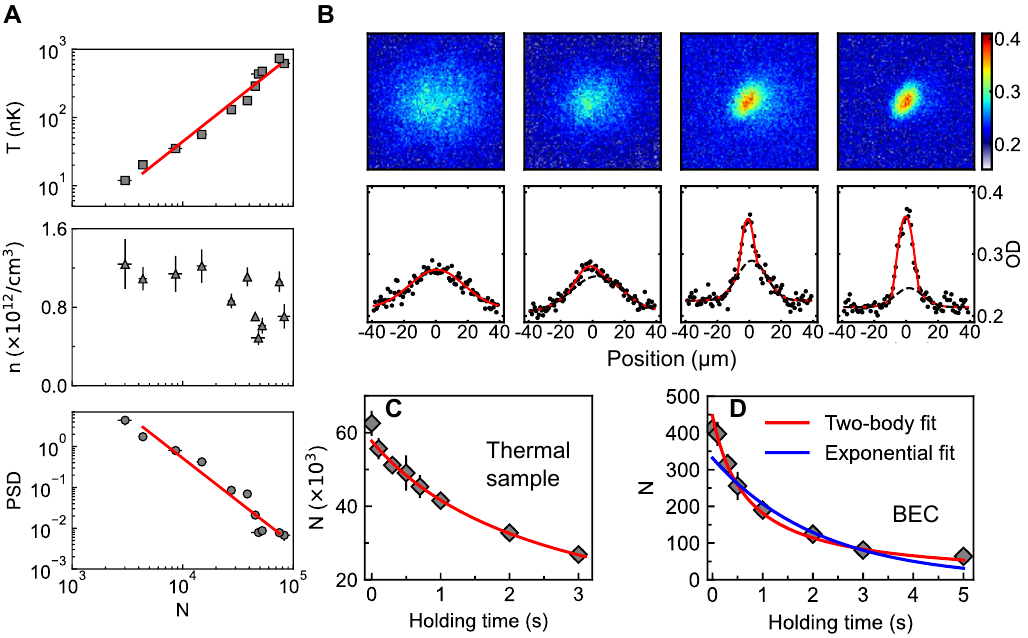} 
	\caption{\textbf{Evaporative cooling and BEC phase transition.}
	(A) From top to bottom: trajectories of the temperature $T$, number density $n$, and PSD during the evaporation of molecules. The solid lines in $T$ and $n$ are linear fits on log-log scales used to extract the evaporation efficiency. (B) TOF absorption images (top) and integrated OD profiles (bottom) across the BEC phase transition. The integration is performed along the tilted long axis, and bi-modal fits are applied to determine the BEC fraction. The expansion time is 13 ms, and each image is averaged over 10 consecutive shots. (C) and (D) are loss measurements of the initial thermal sample and the condensate, respectively. Red and blue solid curves are two-body and exponential fits. Error bars represent one standard deviation.}
	\label{fig2} 
\end{figure*}

In the first experiment, we fix the microwave parameters to $\Omega_{\sigma} = 12.97$ MHz, $\Delta_{\sigma} = 12$ MHz, $\Omega_{\pi} = 11.5$ MHz, and $\Delta_{\pi} = 16$ MHz. Under ideal microwave conditions, these parameters result in an almost complete cancellation of the first-order DDI. However, the misalignment between the two microwaves leads to a residual DDI which modifies the shielding effect. Nevertheless, the measured $\beta_{\rm{in}}$ for the initial thermal sample is approximately $8(1)\times 10^{-13} \,\rm{cm^3/s}$ (Fig.~\ref{fig2}C), already four times lower than that achieved with the single $\sigma^+$ microwave~\cite{lin2023microwave,note1}. Notably, even lower $\beta_{\rm{in}}$ has been observed by changing only $\Delta_{\pi}$ to certain values. A more detailed investigation, with more precisely controlled microwave polarizations, is warranted in future studies. 

The $\beta_{\rm{in}}$ achieved at $\Delta_{\pi} = 16$ MHz is already sufficient to support highly efficient and straightforward evaporative cooling. As shown in Fig.~\ref{fig2}, when the ODT depth is lowered along an optimized trajectory, we observe a steady decrease in $T$ from 734(3) nK to 20.3(6) nK after a factor of 10 loss in $N$, following a slope $d \textrm{ln} T/d \textrm{ln}N$ of 1.3(1). At the same time, the peak density $n$ increases moderately from $7(1)\times 10^{11}$ cm$^{-3}$ to $1.1(1)\times 10^{12}$ cm$^{-3}$. Finally, the PSD improves significantly from 0.008 to 1.7(2), corresponding to an approximately constant and high evaporation efficiency $- d \textrm{ln} PSD / d \ln N$ of 2.1(2). 

The evaporation efficiency remains high during further evaporative cooling. As shown in the averaged absorption images with a TOF of 13 ms in Fig.~\ref{fig2}B, a bi-modal distribution emerges as the molecule number is reduced to around 4000, signaling the onset of the BEC phase transition. Averaging of the absorption images is necessary to improve the signal-to-noise ratio (SNR), as the optical density (OD) is low due to the small particle number and TOF expansion. At the transition point, the sample temperature measured using the thermal component is 20 nK. At the end of the evaporation, we obtain a BEC with approximately 500(20) $^{23}$Na$^{87}$Rb molecules, with a maximum BEC fraction of 70\%. At this stage, the trap frequencies determined from the sloshing motion of the BEC are $2\pi \times [18.5(1.4), 18.5(1.4), 51(2)]$ Hz. 
The typical peak density, determined from \textit{in situ} absorption imaging, is $n = 1.8(5) \times 10^{12}~\textrm{cm}^{-3}$~\cite{note1}. The trap lifetime of the condensate, as obtained from exponential fit to the loss measurement in Fig.~\ref{fig3}D, is a long 2.1(2) s. The loss is dominated by two-body processes, with a measured $\beta_{\rm{in}}$ of $3.5(4)\times 10^{-13} \,\rm{cm^3/s}$, which is much lower than that of the thermal sample.

As the microwave fields remain on during the TOF expansion, the effect of the residual DDI can be clearly seen in the condensate shape. From the two images after the phase transition in Fig.~\ref{fig2}B, we can already observe that the condensed part of the sample is noticeably tilted. We attribute this phenomenon to the electrostriction effect, where molecules tend to accumulate along the attractive DDI direction to minimize the total energy of the system~\cite{note1}. 
In addition, we observe that the centers of the thermal part and the BEC part are slightly displaced from each other. Currently, we do not have a clear explanation for this behavior. However, we conjecture that this is from the combined effects of stronger interactions within the condensate compared to the thermal cloud and unavoidable gradients in the microwave field.

\section*{Tunable dipolar interaction and quantum droplet}

\begin{figure*} 
	\centering
	\includegraphics[width=0.8\linewidth]{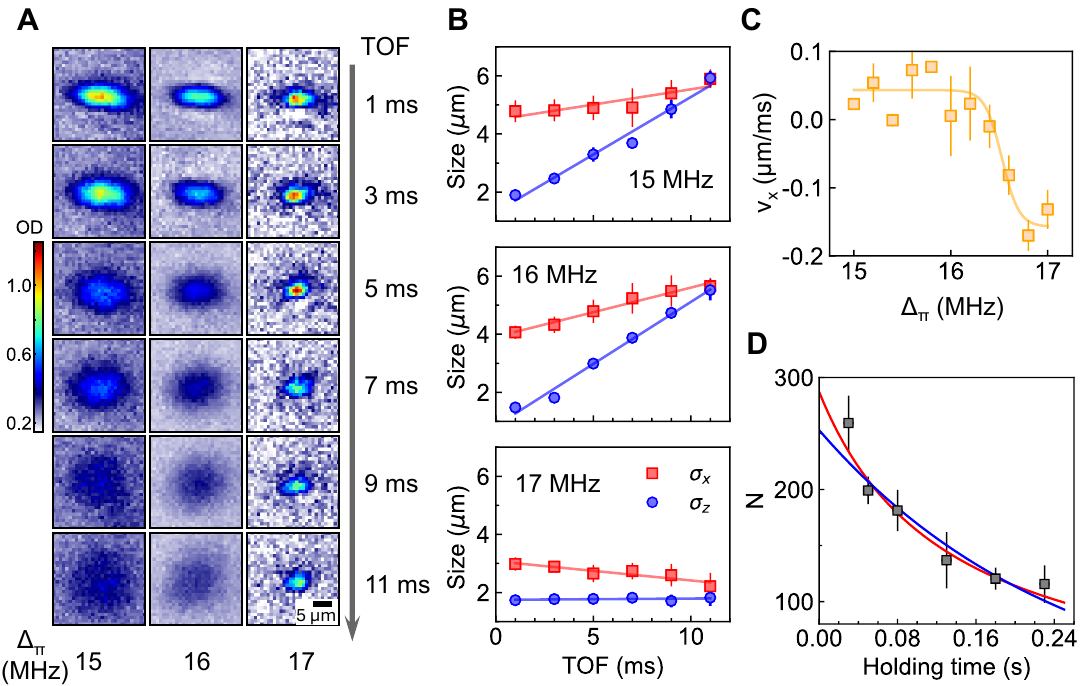} 
	\caption{\textbf{Tunable dipolar interactions.} (A) TOF absorption images with only $\Delta_\pi$ tuned, while all other microwave parameters are fixed, showing the transition between gas-phase BECs ($\Delta_\pi = 15$ MHz and 16 MHz, left and middle) and a self-bound molecular liquid droplet ($\Delta_\pi = 17$ MHz, right). For the gas-phase samples, each image is averaged over 10 consecutive shots. For the droplet, the OD is high enough for single-shot imaging.	(B) Horizontal (red squares) and vertical (blue circles) RMS sizes versus TOF, obtained from 2D fits to the absorption images. Solid lines are from linear fitting. (C) Horizontal expansion velocity $v_x$ as a function of $\Delta_\pi$. The solid curve is a sigmoid fit used to extract the gas-to-droplet transition point~\cite{note1}. (D) Lifetime measurement of the trapped liquid droplet. The red and blue solid curves represent two-body and exponential fits, respectively. Error bars indicate one standard deviation.}
	\label{fig3} 
\end{figure*}

Double microwave dressing allows for highly flexible interaction control by varying different parameters of the two microwave fields. Here, to explore more distinct behaviors of the $^{23}$Na$^{87}$Rb BEC, we focus on tuning the DDI by changing $\Delta_{\pi}$, while keeping the other microwave parameters fixed. Under ideal microwave conditions, adjusting $\Delta_{\pi}$ near the DDI cancellation point is a method to introduce DDI in a controlled manner. Intuitively, as $\Delta_{\pi}$ increases, the contribution from the oscillating dipole becomes smaller, leading to under-compensation of the DDI from rotating dipoles. Conversely, decreasing $\Delta_{\pi}$ results in over-compensation, giving rise to a DDI dominated by oscillating dipoles.

Experimentally, we first prepare the BEC at $\Delta_{\pi} = 16$ MHz and then linearly sweep the microwave frequency over 20 ms to desired $\Delta_{\pi}$. After an additional 5 ms hold time, the sample is released from the trap for TOF imaging. As presented in Fig.~\ref{fig3}, for $\Delta_{\pi} = 16$ MHz or smaller (e.g., 15 MHz), the sample sizes always increase during TOF, indicating that they are all in the gas phase. However, the tilting angle of the images for $\Delta_{\pi} = 16$ MHz is noticeably more obvious than that for $\Delta_{\pi} = 15$ MHz, especially at longer TOF. This can be attributed to the cancellation effect, where the additional contributions to the DDI from rotating dipoles, caused by microwave misalignment, are counteracted by the larger contributions from oscillating dipoles at smaller $\Delta_{\pi}$. Because of this effect, for $\Delta_{\pi} = 15$ MHz, the residual DDI is smaller than that for $\Delta_{\pi} = 16$ MHz, and the electrostriction effect becomes less pronounced.   

For larger $\Delta_{\pi}$, the expansion dynamics changes drastically. For instance, at $\Delta_{\pi} = 17$ MHz, the vertical size $\sigma_z$ remains nearly constant up to 11 ms TOF, although its real value cannot be fully determined due to the limited resolution of our imaging system. While the horizontal size $\sigma_x$ is much larger and resolvable, it actually decreases with TOF, which we attribute to the position dependence of the $\sigma^+$ microwave ellipticity. This can compromise the shielding effect and may also drive the droplet to higher density regimes. Both of these effects can lead to faster losses and sample size shrinking. Notably, although the loss during the free expansion is fast, the trapped droplet has a rather long effective lifetime of 250(36) ms, as presented in Fig.~\ref{fig3}D.

These observed expansion behaviors are consistent with that expected for a self-bound dipolar molecular quantum droplet as predicted for increasingly larger strength of the \textit{anti-dipolar} interaction~\cite{jin2025bose,langen2025dipolar}. A unique feature of the microwave-dressed molecular quantum droplet is that it can be stabilized by the long-range shielding core. Even for the small molecule number used here, the shielding core can still prevent complete collapse and lead to the formation of a droplet.

To determine the transition point between gas-phase BEC and the liquid droplet, we extract the initial horizontal expansion velocity $v_x$ from the TOF measurement. We choose to only use data points at shorter TOF durations in order to minimize the influences of the residual DDI and the ellipticity variation during the TOF. As shown in Fig.~\ref{fig3}C, by fitting $v_x$ with a sigmoid function, we obtain a transition point of $\Delta_{\pi} = 16.5(1)$ MHz. 

Because $\sigma_z$ is too small to be determined experimentally, the density of the droplet cannot be measured directly. Based on the loss measurement in Fig.~\ref{fig3}D and assuming that $\beta_\textrm{in}$ is the same as in the gas-phase BEC sample, we can estimate the density using the two-body loss model. The result suggests that the density of the droplet created with $\Delta_{\pi} = 17$ MHz is an order of magnitude higher than that of the gas phase, reaching $5.9\times 10^{13}$ cm$^{-3}$.

\section*{Discussion}

\begin{figure} 
	\centering
	\includegraphics[width=0.9\linewidth]{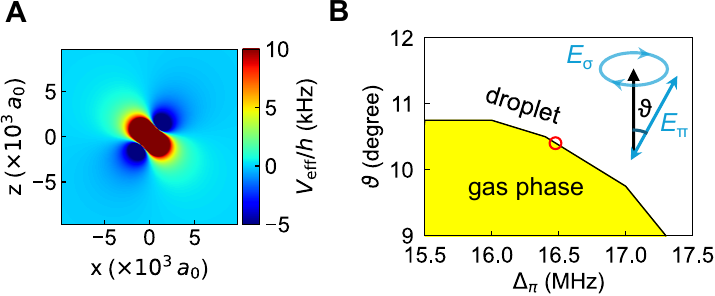} 
	\caption{\textbf{Effect of misalignment between microwaves.} (A) The effective potential in the $xz$ plane illustrates effect of the misalignment, showing the tilted DDI axes. The misalignment is assumed to be in the $xz$ plane with $\vartheta = 10^{\circ}$. (B) The calculated gas-to-droplet phase diagram shows that the transition point $\Delta_\pi$ depends sensitively on $\vartheta$. The red circle marks the experimentally measured gas-to-droplet transition point at $\Delta_\pi = 16.5$ MHz, which can be reproduced for $\vartheta = 10.4^\circ$. Inset: the misalignment angle $\vartheta$ is defined as the angle between the polarization direction of the $\pi$ microwave and the propagation direction of the $\sigma^+$ microwave. }
	\label{fig4} 
\end{figure}

We emphasize that the misalignment between the two microwaves, though not fully characterized, plays a crucial role in enabling the observation of the gas-to-droplet transition within such a small range of $\Delta_\pi$. The effect of the misalignment is illustrated in Fig.~\ref{fig4} through our theoretical modeling. In this calculation, the electric field of the $\pi$ polarization microwave is assumed to be tilted by an angle $\vartheta$, as defined in the inset of Fig.~\ref{fig4}B. Although the actual misalignment in our system may be in an arbitrary azimuthal direction, here we assume it lies in the $xz$ plane. This misalignment causes the interaction potential to tilt correspondingly in the same plane (Fig.~\ref{fig4}A). Especially, the attractive interaction lies in the same overall direction as the tilted long axis of the condensate shown in Fig.~\ref{fig2}B. The observed electrostriction effect can thus be qualitatively explained~\cite{note1}. 

As shown in the gas-to-droplet phase diagram in Fig.~\ref{fig4}B, the gas-to-droplet transition point is sensitive to the degree of misalignment. When $\vartheta = 0$, the droplet phase will happen at much larger $\Delta_\pi$, where loss shielding could become much less efficient. In contrast, the current degree of misalignment, while limited by our ability to control the microwave, allows the transition to occur within an experimentally favorable range, without re-introducing excessive losses. 

The observed transition point at around $\Delta_\pi = 16.5$ MHz can be reproduced by assuming $\vartheta = 10.4^\circ$ in the $xz$ plane. However, this $\vartheta$ is outside the upper bound of the measured value. A more quantitative agreement between theory and experiment could be achieved by adjusting $\vartheta$ and the azimuthal orientation between the two microwaves, and the ellipticity angle of the $\sigma^+$ microwave. Nevertheless, we choose to improve the agreement more rigorously in future work, by refining the characterization and control of the microwave field. We believe that carefully controlled misalignment, much like ellipticity~\cite{chen2023field}, could also serve as a powerful tool for tuning the DDI and uncovering additional phases of the system. 

Using the relatively weak DDI generated by the set of dual microwave dressing parameters explored in this work, we have created both molecular condensates in the gas phase and in the liquid droplet phase. Leveraging the flexible controllability of dual microwave dressing, it should be possible to tune the interaction over a much wider range without incurring significant losses. The high density and relatively long lifetime of the droplet make it a promising starting point for realizing UPMs in optical lattices with high filling factors. The combination of strong DDI and high filling optical lattices offers a promising platform for future studies of dipolar quantum many-body physics with strong long-range interactions.

~\\

\textbf{Acknowledgments:}~We thank Guanghua Chen for her contributions at the early stage of the experiment, Wei Zhang and Xinyuan Hu for insightful discussions on theoretical modeling, and the groups of Xinyu Luo and Sebastian Will for valuable discussions and for sharing many technical details of their experiments. This work is supported by the Innovation Program for Quantum Science and Technology of China (2024ZD0300600), the Hong Kong RGC General Research Fund (Grants 14301620 and 14302722) and the Collaborative Research Fund (Grant No. C4050-23G), National Key Research and Development Program of China (Grant No. 2021YFA0718304), NSFC (Grants No. 12135018), and Guangdong Provincial Quantum Science Strategic Initiative (Grant No. GDZX2303002).

%\bibliography{refs} 

%

%%%%%%%%%%%%%%%%%%%%%%%%%%%%%%%%%%%%%%%%%%%%%%%%%%%%%%%%%%%%%%%%%%%%%%%%%%%%%%%%%%%%%%%%%%%%%%%%%%%%%
%%%%%%%%%%%%%%%%%%%%%%%%%%%%%%%%%%%%%%%%%%%%%%%%%%%%%%%%%%%%%%%%%%%%%%%%%%%%%%%%%%%%%%%%%%%%%%%%%%%%%
%%%%%%%%%%%%%%%%%%%%%%%%%%%%%%%%%%%%%%%%%%%%%%%%%%%%%%%%%%%%%%%%%%%%%%%%%%%%%%%%%%%%%%%%%%%%%%%%%%%%%

%%%%%%%%%%%%%%%%%%%%%%%%%%%%%%%%%%%%%%%%%%%%%%%%%%%%%%%%%%%%%%%%%%%%%%%%%%%%%%%%%%%%%%%%%%%%%%%%%%%%%
%%%%%%%%%%%%%%%%%%%%%%%%%%%%%%%%%%%%%%%%%%%%%%%%%%%%%%%%%%%%%%%%%%%%%%%%%%%%%%%%%%%%%%%%%%%%%%%%%%%%%
%%%%%%%%%%%%%%%%%%%%%%%%%%%%%%%%%%s%%%%%%%%%%%%%%%%%%%%%%%%%%%%%%%%%%%%%%%%%%%%%%%%%%%%%%%%%%%%%%%%%%%

\renewcommand{\thefigure}{S\arabic{figure}}
\renewcommand{\thetable}{S\arabic{table}}
\renewcommand{\theequation}{S\arabic{equation}}
\setcounter{figure}{0}
\setcounter{table}{0}
\setcounter{equation}{0}

\newpage

\section*{Supplementary materials}

\subsection{Optimized sample preparation}

To improve the initial condition for evaporative cooling, the number of ground-state $^{23}\text{Na}^{87}\text{Rb}$ molecules is increased to almost twice that reported in our previous study~\cite{lin2023microwave} by a series of optimization. 

Initially, we tried, with limited success, to improve the atom-to-molecule conversion efficiency in the magneto-association (MA) process, which is less than 10\%, even though the fraction of lost atoms is always larger. In addition, heating is always observed during MA, which raises the temperature of the $^{23}\text{Na}^{87}\text{Rb}$ samples significantly higher than those of the $^{23}\text{Na}$ and $^{87}\text{Rb}$ atomic clouds. Although our results are not conclusive, these issues are likely due to complicated few-body processes during the magnetic field ramping across the atomic Feshbach resonance. For example, the recently observed halo trimer in Bose-Fermi mixtures~\cite{Chuang2025halo} should also be present in the Bose-Bose system used here. The existence of this trimer may directly reduce the fraction of atoms that can be converted into Feshbach molecule. The accompanying three-body processes could also cause rapid loss and heating. 

Given the low association efficiency, we obtain the larger $^{23}\text{Na}^{87}\text{Rb}$ samples by first increasing the number of atoms. By carefully optimizing the evaporative cooling and magnetic field ramping sequence, we are able to produce a BEC of $^{23}\text{Na}$ and a thermal gas of $^{87}\text{Rb}$, each containing approximately $1.5\times 10^6$ atoms as the starting point for association. 

Second, to reduce molecule loss from collisions with high-density residual atoms, we shorten the duration of MA and atom removal processes. To this end, we set the final magnetic field to 346.4 G, which is much closer to the 347.65 G heteronuclear atomic Feshbach resonance than the 335 G final magnetic field used in our earlier works~\cite{Guo2016}. Since the binding energy of the Feshbach molecules is significantly smaller at this field, the coupling strength of the pump transition of the STIRAP is much weaker. Fortunately, after optimizing the Raman laser beams, we are able to achieve a STIRAP efficiency of 91\% with a pulse duration of 60~$\mu$s. The removal of residual $\text{Na}$ and $\text{Rb}$ atoms is performed after the STIRAP using 40~$\mu$s light pulses, which drive the same optical pumping and cycling transitions used for high-field imaging~\cite{jia2020detection}. 

Finally, to suppress rapid loss from non-shielded molecule-molecule collisions, the dual microwave shielding is ramped on in 100~$\mu$s immediately after the STIRAP, simultaneously with the atom removal process.

\subsection{Microwave generation and calibration}

\begin{figure*} 
	\centering
	\includegraphics[width=0.85\linewidth]{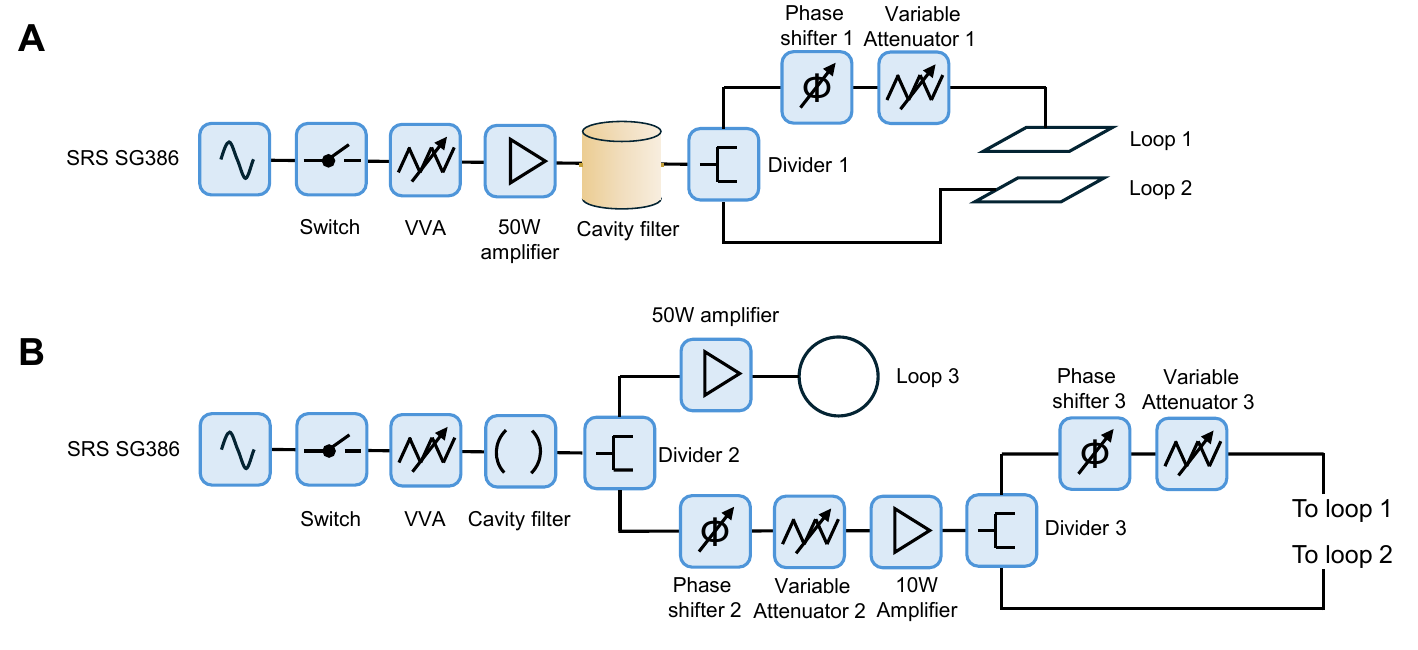} 
	\caption{\textbf{Schematic of the microwave system.} (A) Generation of the $\sigma^+$ microwave. The low-noise, high-power signal is fed into two loop antennas (Loop 1 and Loop 2) for polarization control. (B) Generation of the $\pi$ microwave. The main branch is fed into a single loop (Loop 3) to generate the primary $\pi$ microwave. The secondary branch is combined with the pair of loop antennas (Loop 1 and Loop 2) to cancel the projection component caused by the imperfect polarization of the main branch.}
	\label{figS1}
\end{figure*}

As illustrated in Fig.~\ref{figS1}A, the $\sigma^+$ microwave is generated using a SRS SG386 signal generator. After passing through a switch (Mini-Circuits ZASWA-2-50DR+) and a voltage-variable attenuator (General Microwave D1954) for power control, the signal is amplified to nearly 50 W by a low-noise amplifier (Microwave Amps AM43). The high-power signal is then filtered by a home-made cylindrical microwave cavity with bandwidth less than 1 MHz to suppress phase noise before being split into two branches by a power divider. They are then fed into the crossed full-wavelength loop antennas placed on the top and bottom of the glass vacuum cell. With their feeding points offset by $90^\circ$, the antennas emit linearly polarized microwaves with perpendicular polarization directions. By tuning the relative phase and amplitude between the two signals using a phase shifter (Phase Shifter 1) and an attenuator (Attenuator 1) placed in one of the branches, their superposition produces a nearly pure $\sigma^+$ polarization microwave field at the center of the ODT.  

The $\pi$ microwave, shown in Fig.~\ref{figS1}B, is also generated by an SG386 signal generator, followed by a switch and a voltage-variable attenuator. This low-power signal is filtered by two bandpass cavity filters and then split into two branches. The first branch is amplified using another 50 W power amplifier and fed into a third loop antenna (Loop 3) placed orthogonal to the two $\sigma^+$ antennas to generate the main $\pi$ microwave signal. The purity of the polarization is first optimized by adjusting the orientation of the loop. The residual circularly polarized components are then canceled by feeding the second branch into the two $\sigma^+$ antennas after amplifying it with a 10 W amplifier (RF Bay, JPA-2000-6500-10) and splitting it into two sub-branches. Before being amplified, the second branch signal passes through a phase shifter (Phase Shifter 2) and an attenuator (Attenuator 2) to obtain the amplitude and phase required for the best polarization impurity cancellation. Arbitrary polarization control is achieved using an additional phase shifter (Phase Shifter 3) and attenuator (Attenuator 3) placed in one of the sub-branches. 

The microwave polarizations are calibrated using the method described in our previous work at low microwave power~\cite{lin2023microwave}, where the quantization axis is defined by the magnetic field. For the $\sigma^+$ microwave, the measured polarization ratios are $E_{\sigma^+} : E_{\sigma^-} : E_{\pi} = 1 : 0.03 : 0.08$, while the polarization ratios for the $\pi$ polarization signal are $E_{\sigma^+} : E_{\sigma^-} : E_{\pi} = 0.04 : 0.05 : 1$. Based on these ratios, we estimate that the propagation direction of the $\sigma^+$ microwave lies between $6.1^\circ$ and $9.5^\circ$ relative to the quantization axis, while the direction of the electric field of the $\pi$ microwave lies between $0.4^\circ$ and $3.4^\circ$ relative to the same axis. Unfortunately, the azimuthal orientations of both the individual polarizations and their relative alignment cannot be determined.

The situation becomes more complicated at high power, where the quantization axis becomes largely defined by the electric field of the microwave. Assuming the polarizations of the microwave signals in their own respective reference frames are unchanged with power levels, we can estimate the lower and upper bounds of the misalignment angle between the two microwaves as $2.7^\circ \leq \vartheta \leq 9.9^\circ$.  This large uncertainty is the primary limiting factor in achieving full cancellation of the first-order DDI and establishing a quantitative agreement between theory and experiment. While the uncertainty in the azimuthal orientation between the microwaves does not affect the gas-to-droplet phase diagram, (if we ignore the small ellipticity of the $\sigma^+$ microwave), it influences the observed absorption images, as the interaction effect is projected onto the lab frame during detection. Developing more precise control of the polarizations and their misalignment will be a focus of our future work.

To measure the Rabi frequencies at high power, we use the dressed-state spectroscopy. The maximum attainable Rabi frequency for the $\sigma^+$ polarization microwave is $\Omega_{\sigma} = 25~\text{MHz}$, while the maximum $\Omega_{\pi}$ is measured to be 11.5 MHz. At these high Rabi frequencies, the phase noise of the microwave signals for sidebands near the Rabi frequencies must be reduced to below the -160 dBc/Hz level to suppress one-body losses. With our highly filtered microwave system, we routinely obtain a one-body lifetime of 12(2) s for Rabi frequencies larger than 10 MHz. From this long one-body lifetime, we can infer that the microwave phase noises are indeed below -160 dBc/Hz. We believe that this could be further improved by adding another microwave resonator after the amplifier in the $\pi$ polarization branch. However, this is not currently implemented due to the -3 dB power attenuation in the cavity transmission, which would lower $\Omega_{\pi}$ to below 10 MHz.

\subsection{Detection of the molecular BEC}

Due to the absence of cycling transitions, ground-state $^{23}\text{Na}^{87}\text{Rb}$ molecules need to be dissociated into $^{23}\text{Na}$ and $^{87}\text{Rb}$ atoms for detection. This process starts from ramping off the dual microwave fields in 50~$\mu$s, transferring the dressed molecules to the bare $| J=0, m_J=0 \rangle$ state. Traditionally, the dissociation is performed in two steps. First, the ground-state molecules are converted back to Feshbach molecules with a reversed STIRAP. The Feshbach molecules are then dissociated either by magneto-dissociation (MD), using a reversed ramp of the magnetic field across the Feshbach resonance, or by single-photon photo-dissociation (PD) via a selected excited atomic state~\cite{jia2020detection}. However, to obtain accurate density distribution of the molecular sample, the whole dissociation process must occur after the TOF and immediately before imaging using the atomic transition.

As discussed in the main text, at the end of the evaporation, the density of the gas-phase BEC is over $10^{12}~\rm{cm^{-3}}$, while the density of the droplet reaches the $10^{13}~\rm{cm^{-3}}$ level. Without microwave shielding, $\beta_{\rm{in}}$ is on the $10^{-10}\,\rm{cm^3/s}$ level, and the two-body collision limited lifetime of these samples would be less than 1 ms. Unfortunately, the rapid two-body loss is expected to persist during the TOF until the density decreases to very low levels. Since the number of molecules in the condensate is small, we are already relying on the averaging of multiple absorption images to improve the SNR. Losses during detection would further reduce the SNR, making it even more challenging. 

A seemingly straightforward solution to mitigate the loss, as implemented in this work, is to keep the microwave fields on during the TOF expansion. However, since the microwave polarization is only optimized at the trap center, polarization impurities may arise during free fall from the position-dependent relative phases and amplitudes of the microwave fields. Notably, at 10 ms and 20 ms TOF, the sample drops about 0.5 mm and 2 mm from the trap center, respectively. For our antenna configuration, we estimate that a 1 mm displacement from the trap center will introduce an ellipticity angle of approximately $5^{\circ}$ relative to the polarization at the trap center. In future work, this problem will be addressed by dynamically tuning the microwave signals during the TOF or by employing a different antenna design.  

In our system, the diameters of the horizontally propagating pump and dump Raman laser beams are 110~$\mu$m $\times$ 110~$\mu$m and 250~$\mu$m $\times$ 250~$\mu$m, respectively. Thus, the sample will be out of the Raman beams after only 5 ms of TOF. To track the sample for longer TOF durations, we use a galvo (LBTEK, EM-RP60) to dynamically steer the Raman beams vertically. The rotating angle of the galvo is calibrated for each TOF using a thermal sample and optimized to ensure consistent detected molecular numbers across different TOF durations. With this setup, we can probe the molecular sample for up to 20 ms of TOF.

Another challenge is managing the various heating effects during the whole detection process. Because of the small molecule number, the BEC phase transition can only occur at temperatures below 10 nK, making the system especially sensitive to heating. With the microwave shielding on, the two-body loss is sufficiently slow, rendering heating from the ``anti-evaporation'' effect during TOF negligible. However, losses and heating during dissociation and imaging, which occur after the microwave shielding is turned off, can still have significant impacts. The reversed STIRAP is fast, with negligible heating. In contrast, MD of Feshbach molecules must be slow, typically on the order of 1 ms, as the dissociation efficiency and the amount of heating are both determined by the ramping speed of the magnetic field. Since Feshbach molecules also experience two-body losses at a rate comparable to that of bare ground-state molecules, the losses and heating during the MD process are significant. For single-photon PD, although the process is very fast, requires only tens of $\mu$s, the amount of heating reaches at least hundreds of $\mu$K due to the broad atomic transition linewidth~\cite{jia2020detection}. This level of heating is sufficient to completely alter the density distribution of the condensate during the 40 $\mu$s imaging time. 

To achieve fast dissociation with minimum heating, we change the dissociation process to a single step, directly transferring ground-state molecules to free atoms using STIRAP. For this purpose, we first step the magnetic field to 347.7 G, which is 50 mG above the Feshbach resonance, while the molecules are still in the ground state. The pump laser frequency is also tuned to maximize the two-photon bound-to-free transfer efficiency. After optimization, this STIRAP PD process can robustly achieve 70$\%$ efficiency within 60 $\mu$s. The measured heating is less than 1~$\mu$K, and its effect on the density distribution during the 40 $\mu$s imaging time is barely detectable. While the source of the 1~$\mu$K heating has not been studied in detail, the most likely cause is the excess photon energy above the dissociation limit. 

Finally, we detect the distribution of the molecular cloud using a high-magnetic-field imaging scheme~\cite{jia2020detection} on the fragmented $^{87}\text{Rb}$ atoms. We use $^{87}\text{Rb}$ instead of $^{23}\text{Na}$ because of its larger absorption cross section and smaller photon recoil. These properties are important for minimizing heating during imaging and for maximizing the SNR. 

The objective of our imaging system has a numerical aperture (N.A.) of 0.2. The imaging resolution ($1/\sqrt{e}$ Gaussian width), measured with our smallest molecular droplet, is $\sigma_{\rm{limit}}=1.3~\mu \rm{m}$. This is just enough to resolve the horizontal size $\sigma_x$ of the gas-phase BEC and the droplet, and the vertical size $\sigma_z$ of the gas-phase BEC after TOF expansion. For both the the droplet and the trapped gas-phase BEC, the measured $\sigma_z$ is very close to $\sigma_{\rm{limit}}$. This makes it difficult to determine the real value of $\sigma_z$.

\subsection{Loss measurement and condensate lifetime}

\begin{figure} 
	\centering
	\includegraphics[width=0.95\linewidth]{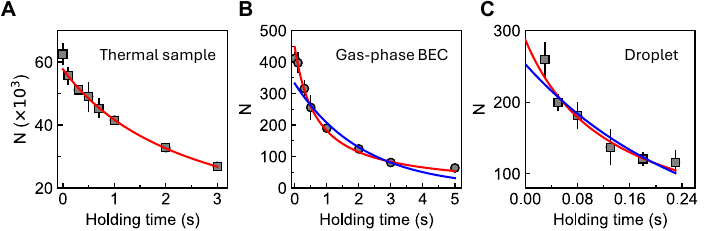} 
	\caption{\textbf{Loss and lifetime measurement.} (A) Time evolution of the number of molecules $N$ for a 730 nK thermal sample. (B) The same measurement for a trapped gas-phase BEC, and (C) for a trapped liquid droplet. The solid curves represent exponential fits (blue), used to extract the effective lifetime, and two-body fits using Eq.~\ref{eqS1} (red). Error bars represent one standard deviation.}
	\label{figS2} 
\end{figure}

To determine the trap lifetime of the gas-phase BEC, we hold the sample in the final ODT for various durations and measure the evolution of molecule number $N$ for $\Delta_{\pi} = 16$ MHz, as shown in Fig.~\ref{figS2}B. From an exponential fit (blue curve), the $1/e$ lifetime is $\tau = 2.1(2)~\rm{s}$. This is much shorter than the 12~s one-body lifetime measured with thermal gases. However, the loss of molecules fits much better to the two-body loss model (red curve)

\begin{equation}
	N(t) = \frac{N_0}{1 + N_0 B t },
	\label{eqS1}
\end{equation}

with $N_0$ the initial molecule number and $B$ the loss rate in unit of $1/\rm{s}$. This indicates that the loss is dominated by two-body inelastic collisions and $\tau$ is just an effective lifetime. 

However, the extraction of $\beta_{\rm{in}}$ for the BEC sample is complicated by additional loss from free evaporation as a result of the shallow final trap depth. Indeed, we observe that the condensate fraction remains large even after more than 2~s of holding time, during which near 70\% of the molecules are lost. This behavior is consistent with free evaporation occurring during the loss measurement. Nevertheless, for the same set of data, if we ignore loss from free evaporation and use the peak densities measured \textit{in situ} at each holding time, we obtain a $\beta_{\rm{in}}$ of $3.5(4)\times 10^{-13}\,\rm{cm^3/s}$ by fitting the density evolution to $n(t) = n_0/(1+n_0 \beta_{\rm{in}}t)$.

Because of the residual DDI during TOF, deducing the trapped density distribution from the TOF data is complicated. As a result, we use \textit{in situ} absorption images to measure the density. Given the limited resolution of our imagine system, the relatively larger $\sigma_{x,{\text{measure}}}$ is deconvoluted with $\sigma_{\text{limit}}$ using $\sqrt{\sigma_{x,{\text{measure}}}^2 - \sigma_{\text{limit}}^2}$ to determine the actual $\sigma_x$. For $\sigma_z$, however, the measured value is not significantly larger than $\sigma_{\text{limit}}$, making deconvolution unreliable. Instead, we estimate $\sigma_z$ using trap frequency scaling as $\sigma_z = \sigma_x \times \omega_x / \omega_z$. This is a reasonable approximation for gas-phase BECs with weak DDI. Once the sizes are determined, the peak density $n$ at each holding time can be calculated using a Gaussian distribution with cylindrical symmetry as $N / [(2\pi)^{3/2} \sigma_x^2 \sigma_z]$. The molecule number $N$ can always be measured accurately, even with limited resolutions. 

For completeness, in Fig.~\ref{figS2}A we also present the loss measurement at the same microwave parameters for a thermal sample with an initial temperature of 730 nK. The sample temperature is nearly constant during the several seconds holding time. The $\beta_{\rm{in}}$ extracted from this measurement, following the method used in our previous work~\cite{lin2023microwave}, is $8(1)\times 10^{-13}\,\rm{cm^3/s}$.

\subsection{Density and size of the droplet}   

The in-trap lifetime of the quantum droplet created with $\Delta_{\pi} = 17$ MHz is measured using the same method, as depicted in Fig.~\ref{figS2}C. The measured $1/e$ lifetime is a relatively long 250(36)~ms. However, significant losses are also observed during the less than 20 ms TOF, which we attribute to variations in the microwave polarization at different positions during the TOF. These variations can reduce the effectiveness of loss suppression and drive the droplet into a higher density regime, both of which can result in faster losses. 

As the density distribution of the droplet is strongly modified by the DDI, the vertical size $\sigma_z$ cannot be determined using the trap frequency scaling method. Thus, the density and $\beta_{\rm{in}}$ of the droplet cannot be measured using the sizes. To estimate the density, we fit the droplet loss data with Eq.~\ref{eqS1} and extract a loss rate of $B = 2.6(4)\times 10^{-2}/\rm{s}$. Assuming a constant effective trapping volume and the same $\beta_{\rm{in}}$ as for the gas phase BEC at $\Delta_{\pi} = 16$ MHz, the mean density of the trapped droplet can be estimated from the relation $ \left\langle n \right\rangle = N_0 (B/\beta_{\rm{in}})$. With this simplified calculation, we obtain a mean droplet density of $2.1\times 10^{13}\,\rm{cm^{-3}}$. If we assume the density distribution is Gaussian, the peak density will be $2\sqrt{2} \left\langle n \right\rangle \approx 5.9 \times 10^{13}\,\rm{cm^{-3}}$, which is much higher than that of the gas-phase BEC.  

The vertical size $\sigma_z$ can also be estimated from the density. As the trapping potential is cylindrical symmetric, we approximate the density distribution using a 3D Gaussian as $N_0/[(2\pi)^{3/2}\sigma_x^2\sigma_z]$, where $N_0$ and $\sigma_x$ are both measured. From this, we estimate $\sigma_z$ to be approximately 42 nm. This is indeed much smaller than our imaging resolution, indicating that the observed $\sigma_z$ is primarily determined by the resolution limit.

The OD of the trapped droplet created with $\Delta_{\pi} = 17$ MHz is about 1, which is just slightly higher than the OD of 0.8 observed for the trapped gas-phase BEC created with $\Delta_{\pi} = 16$ MHz. However, since the droplet is self-bound during TOF, even with molecule losses, the SNR from single-shot images is still enough for the measurement. Nevertheless, the OD calculated from the estimated high density is over 100, which is significantly higher than the observed value. This discrepancy can be explained by the convolution of the small $\sigma_z$ with the imaging resolution $\sigma_{\rm{limit}}$. Since the molecule number is conserved and $\sigma_{\rm{limit}} \gg \sigma_z$, the observable OD is reduced by a factor of $\sigma_z/\sigma_{\rm{limit}}$, which is approximately 30. As a result, the observable OD is reduced to about 3.8, which is much closer to the measured value.

Using the effective potential~\cite{deng2025two}, we calculate that the $C_6$ coefficient along the z-direction ($\theta = 0$) for $\Delta_{\pi} = 17~\rm{MHz}$ is $3.6 \times 10^7$ atomic units. The characteristic length scale, i.e., the size of the shielding core, is determined as $(m C_6/\hbar^2)^{1/4} = 1638a_0$ or 87 nm. The $\sigma_z$ estimated above is therefore smaller than the size of the shielding core. This suggests that the droplet created with $\Delta_{\pi} = 17$ MHz is most likely already in a 2D monolayer configuration~\cite{Ciardi2025mono}.

\subsection{Gas-to-droplet transition}   

\begin{figure}[ptb]
	\centering
	\includegraphics[width=1\linewidth]{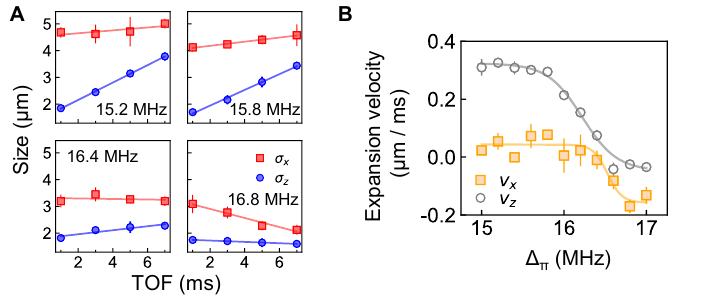}
	\caption{\textbf{Gas-to-droplet transition.} (A) Example TOF expansion data for several $\Delta_\pi$. The solid lines are linear fits for extracting the expansion velocities. (B) The horizontal and vertical expansion velocities $v_x$ and $v_z$. The solid curves are sigmoid fits for determining the transition point. }
	\label{figS3}
\end{figure}

As illustrated in Fig.~\ref{fig3}B and Fig.~\ref{figS3}A, the expansion behavior changes dramatically near the gas-to-droplet phase transition. To determine the $\Delta_\pi$ at which the gas-to-droplet transition happens, we examine the change in expansion velocity. A complication in measuring the expansion velocity is the residual DDI during TOF, which can alter the energy of the sample and, consequently,  the expansion. In additional, as discussed previously, the ellipticity angle of the $\sigma^+$ microwave changes during the TOF. To partially mitigate these issues, we use only data points from smaller TOF durations. Furthermore, for simplicity, we empirically apply a linear fit to extract the expansion velocity.

Figure~\ref{figS3}B shows the measured horizontal ($v_x$) and vertical ($v_z$) expansion velocities. To extract the transition point, we fit them to a sigmoid function with an offset:
\begin{equation}
	v_{x,z} = \frac{A}{1 + e^{(f-f_0)/\delta f}} + A_0,
	\label{eqS2}
\end{equation}

where $f_0$ is $\Delta_\pi$ at the transition point,  $\delta f$ is the width, and $A$ is the amplitude, and $A_0$ is the offset. From the fit to $v_x$, the transition point is determined to be $\Delta_\pi = 16.5(1) $ MHz, while for $v_z$, it is $\Delta_\pi = 16.34(4)$ MHz. As the vertical size $\sigma_z$ is very close to the imaging resolution for both the gas-phase data at 1 ms TOF and the self-bound droplet, only $v_x$ and the transition point obtained from $v_x$ is presented in the main text (Fig.~\ref{fig3}C). 

The horizontal expansion velocity $v_x$ becomes negative for the droplet. As discussed in the main text, this is attributed to the position dependent ellipticity, which could drive the droplet to higher density regime or compromise the loss shielding and thus number losses. Both of these effect could lead to smaller sample sizes at longer TOF and thus the negative $v_x$.

\subsection{Theoretic modeling}

The $\sigma^+$ polarization microwave induces a rotating dipole with an effective dipole moment $d_{\rm{eff}} = d_0/[12(1 + \Delta_{\sigma}^2/\Omega_{\sigma}^2)]^{1/2}$. The first-order DDI between rotating dipoles separated by a distance $r$, $\propto d_{\rm{eff}}^2 (3 \cos^2\theta - 1)/r^3$, has an ``anti-dipolar'' form: it is repulsive for $\theta = 0^{\circ}$ and attractive for $\theta = 90^{\circ}$, where $\theta$ is the polar angle defined relative to the quantization axis. The microwave dressing also modifies the second-order DDI, the rotational van der Waals interaction ($\propto 1/r^6$), and makes it anisotropic. The combined effects of the first-order and second-order DDIs lead to a long-range barrier in the dressed state, preventing colliding molecules from reaching short range, thereby reducing $\beta_{\rm{in}}$ by suppressing the formation of two-molecule complexes. However, the attractive component of the first-order DDI cannot be fully compensated by the second-order DDI. As a result,  for $\theta > 54.7^{\circ}$ a shallow potential well forms at an even longer range than the barrier. For certain values of $\theta$, such as $90^{\circ}$, this potential well is sufficiently deep to support field-linked states of molecules. The existence of these field-linked states facilitates significant three-body losses, thereby limiting the extent to which loss suppression can be achieved.

\begin{figure*}[ptb]
	\centering
	\includegraphics[width=0.9\linewidth]{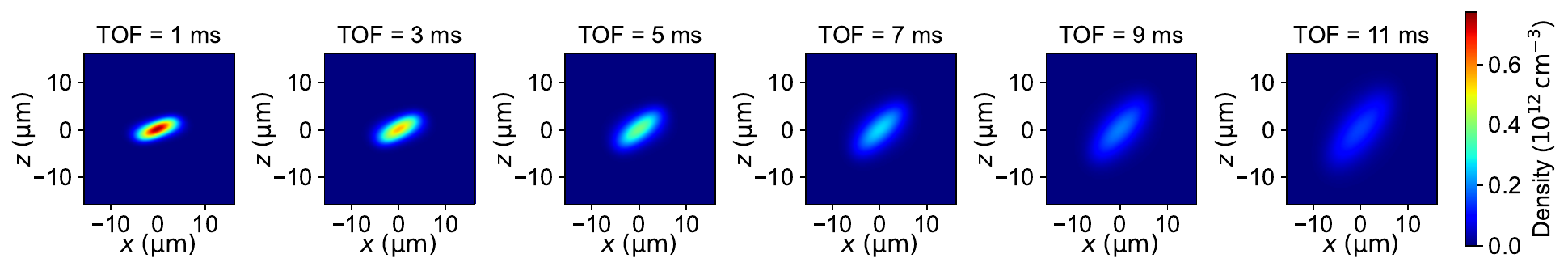}
	\caption{\textbf{Simulation of TOF expansion with microwave misalignment.} The microwave parameters are $\Omega_\sigma= 12.97$~MHz, $\Delta_\sigma= 12$~MHz, $\Omega_\pi= 11.5$~MHz, $\Omega_\sigma= 16$~MHz, $\vartheta=10^\circ$, and $\varphi=0$.}
	\label{figS4}
\end{figure*}

The second microwave with $\pi$ polarization induces an oscillating dipole with an effective dipole moment given by $d_{\rm{eff}} = d_0/[6(1 + \Delta_{\pi}^2/\Omega_{\pi}^2)]^{1/2}$. The DDI between these oscillating dipoles has the opposite sign compared to that between rotating dipoles induced by a $\sigma^+$ polarization microwave. Because of this, with dual microwave dressing, the interaction potential for all $\theta$ can be finely tuned by controlling the contributions from the rotating and the oscillating dipoles via the microwave parameters.  

In the ideal dual-microwave configuration, i.e., $\vartheta=0$, the effective interaction describing molecular collisions has been derived~\cite{deng2025two}:  
\begin{align}
	V_{\rm{eff}}(\textbf{r}) & = \frac{C_6}{r^6}\left[ \sin^4\theta + \frac{w_1}{w_2} \sin^2\theta \cos^2\theta \right.\nonumber\\ 
	&\left. + \frac{w_0}{w_2}(3\cos^2\theta - 1)^2 \right] + \frac{C_3}{r^3}(3\cos^2\theta - 1),
	\label{eqS3} 
\end{align}
where the $C_3$ term and $C_6$ term represent the first-order and second-order DDIs, respectively. A simplified form of $V_{\rm{eff}}$ is presented in Eq.~\ref{eq1} of the main text. The coefficients $w_{m = 0,1,2}$ quantify the relative strengths of different angular interaction components (corresponding to spherical harmonics $|Y_{2m}(\hat{r})|^2$) and can be controlled by varying the microwave parameters. While the $C_6$ term is always repulsive, since
\begin{equation}
	C_3 =\frac{d_0^2}{24\pi\varepsilon_0}(3\cos 2\beta - 1)\cos^2\alpha \sin^2\alpha,
	\label{eqS4} 
\end{equation}  
the first-order DDI can be either repulsive or attractive when the Euler angles $\beta$ and $\alpha$ are tuned by the microwave parameters. In particular, it becomes zero when $\beta$, which controls the contributions from the two excited states, is set to $35.3^\circ$. In this case, the molecules experience a purely long-range repulsive shielding core. Since no field-linked states can form, three-body losses are fully suppressed. Thus, $\beta_{\rm{in}}$ can be further reduced. 

However, in the current experiment, the misalignment between the two microwaves, i.e., $\vartheta\neq0$, breaks the cylindrical symmetry along the $z$-axis and the reflection symmetry of the $xy$ plane. Following a similar formalism as in Ref.~\cite{deng2025two}, we derive the long-range DDI in this case using Floquet theory:

\begin{align}\label{tiled-DDI}
	V_d(\mathbf  r)&=\frac{1}{r^3}\left[C_{3,0}(3\cos^2\theta-1)+\sqrt\frac{2\pi}{15}C_{3,1}Y_{2,-1}(\theta,\phi)\right.\nonumber\\
	&\left.+\sqrt\frac{2\pi}{15}C_{3,-1}Y_{2,1}(\theta,\phi) +\sqrt\frac{8\pi}{15}C_{3,2}Y_{2,-2}(\theta,\phi)\right.\nonumber\\
	&\left.+\sqrt\frac{8\pi}{15}C_{3,-2}Y_{2,2}(\theta,\phi)\right],
\end{align}

where the coefficients $C_{3,m}$ depend on the polar angle $\vartheta$ and azimuth angle $\varphi$ of the $\pi$ polarization microwave field relative to the propagation direction of the $\sigma^+$ polarization field. Here, non-zero $C_{3,\pm1}$ and $C_{3,\pm2}$ reflect the breaking of cylindrical and reflection symmetries. Besides, the coefficients satisfy $C_{3,m}=(-1)^mC_{3,-m}^*$, ensuring $V_d(\mathbf r)$ being real-valued. Unlike the ideal case $\vartheta=0$, the interaction minimum shifts to $(\theta_m,\varphi)$, with $\theta_m=\arctan(e^{-i\varphi}C_{3,1}/(3C_{3,0}-e^{-2i\varphi}C_{3,2}))/2$, as shown in Fig.~4B of the main text.

As a simplified approach to determine the phase diagram in the presence of misalignment, we use the Gross-Pitaevskii (G-P) equation. Specifically, we first calculate the scattering length $a_s$ using a multi-channel scattering calculation and then incorporate the contact-interaction term, $V_\mathrm{contact} = 4\pi\hbar^2 a_s \delta(\mathbf{r})/m$, into the G-P equation. Solving the G-P equation with $V=V_d+V_\mathrm{contct}$, we obtain the phase diagram in Fig.~\ref{fig4}B of the main text.

Additionally, to simulate the TOF experiment, we study the free expansion dynamics by solving the G-P equation in real time. As shown in Fig.~\ref{figS4}, the theoretical results qualitatively agree with the experimental data in Fig.~\ref{fig3}A of the main text. In the presence of the trap, the condensate forms a pancake shape due to the stronger axial trap frequency $\omega_z>\omega_x\approx \omega_y$. However, the condensate is tilted towards the most attractive angle $\theta_m$ of DDI in the $xz$ plane. After releasing the trap, the sample expands during TOF, and the pancake plane simultaneously rotates towards the most attractive direction along the angle $(\theta_m,\varphi)$. This is a clear manifestation of the electrostriction effect, as observed in the experiment.

\end{document}